\documentclass[aps,prd,superscriptaddress,showpacs,preprintnumbers,showkeys,tightenlines,byrevtex,12pt]{revtex4-2}

\usepackage{amsfonts,amssymb,amsmath}
\usepackage{graphicx}
\usepackage{psfrag}
\usepackage{epsfig}
\usepackage{xcolor}
\usepackage[pdfa]{hyperref}
\usepackage[all]{hypcap}
\usepackage{orcidlink}

\hypersetup{
    colorlinks=true,
    linktoc=all,
    menucolor=black,
    citecolor=blue,
    urlcolor=blue
}

\graphicspath{{fig/}}

\begin{document}

\title{\quad\\[1.0cm] Search for a heavy neutrino in $\tau$ decays at Belle}
\noaffiliation
  \author{D.~Liventsev\,\orcidlink{0000-0003-3416-0056}} 
  \author{I.~Adachi\,\orcidlink{0000-0003-2287-0173}} 
  \author{H.~Aihara\,\orcidlink{0000-0002-1907-5964}} 
  \author{S.~Al~Said\,\orcidlink{0000-0002-4895-3869}} 
  \author{D.~M.~Asner\,\orcidlink{0000-0002-1586-5790}} 
  \author{H.~Atmacan\,\orcidlink{0000-0003-2435-501X}} 
  \author{R.~Ayad\,\orcidlink{0000-0003-3466-9290}} 
  \author{V.~Babu\,\orcidlink{0000-0003-0419-6912}} 
  \author{Sw.~Banerjee\,\orcidlink{0000-0001-8852-2409}} 
  \author{M.~Bauer\,\orcidlink{0000-0002-0953-7387}} 
% ...
\collaboration{The Belle Collaboration}

\date{\today}
\begin{abstract}
  We report on a search for a heavy Majorana neutrino in the decays
  $\tau^- \to \pi^-\nu_h$, $\nu_h \to \pi^\pm\ell^\mp$,
  $\ell=e,\mu$. The results are obtained using the full data sample of
  $988\,\mathrm{fb}^{-1}$ collected with the Belle detector at the KEKB
  asymmetric energy $e^+ e^-$ collider, which contains $912 \times 10^6$
  $\tau\tau$ pairs. We observe no significant signal and set 95\%~CL
  upper limits on the couplings of the heavy right-handed neutrinos to
  the conventional SM left-handed neutrinos in the mass range
  $0.2-1.6\,\textrm{GeV}/c^2$. This is the first study of a mixed
  couplings of heavy neutrinos to $\tau$ leptons and light-flavor
  leptons.
\end{abstract}


\maketitle

\hypersetup{
    linkcolor=blue
}


In the Standard model (SM), neutrinos are strictly massless since there
are no right-handed neutrino components. However, experimental data on
neutrino oscillations conclusively show that neutrinos are
massive~\cite{pdg}, though neutrino mass measurements show that their
masses are very small~\cite{katrin}. One approach to resolve this
disagreement is to include right-handed neutrinos, also known as sterile
neutrinos, heavy neutrinos, or heavy neutral leptons (HNL), into the
model. Such particles do not participate in any of the weak, strong, and
electromagnetic interactions; if we exclude gravitation, the only way
they interact with matter is via mixing with left-handed
neutrinos. Singlet right-handed neutrinos may also have Majorana mass,
naturally explaining the smallness of the observed neutrino masses via
the so-called ``see-saw'' mechanism~\cite{seesaw}. One example of the
models realizing such a mechanism is $\nu$MSM~\cite{numsm}. It
introduces three right-handed singlet HNLs, so that every left-handed
particle gets its right-handed counterpart, and manages to explain
neutrino oscillations, dark matter existence, and baryogenesis with the
same set of parameters. HNLs also appear in other extensions of the SM;
see a review in Ref.~\cite{review}.

\begin{figure}[htbp]
  \begin{tabular}{c}
    \includegraphics[width=0.48\textwidth,clip=true]{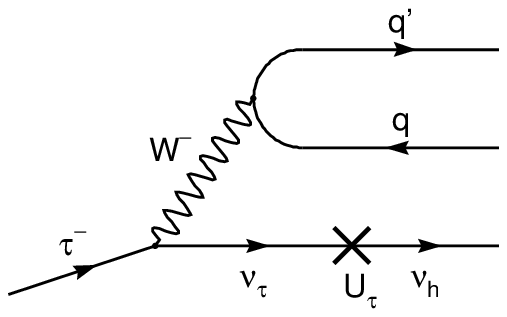}\\
    \includegraphics[width=0.48\textwidth,clip=true]{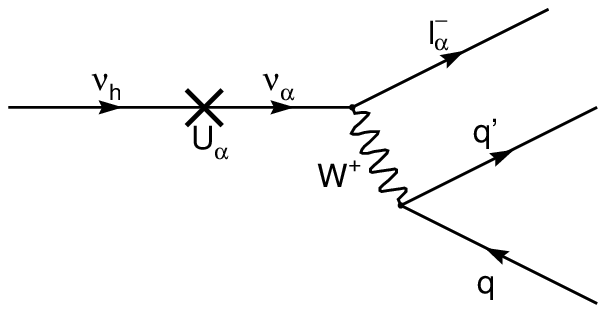}\\
  \end{tabular}
  \caption{Feynman diagrams for HNL production \textit{(top)} and decay \textit{(bottom)}.}
  \label{p:feyn}
\end{figure}

In general, neutrino flavor eigenstates need not to coincide with the
mass eigenstates but may be related through a unitary transformation,
similar to that in the quark sector:
\begin{equation}
\nu_\alpha = \sum_i{U_{\alpha i}\nu_i}, \quad \alpha = e, \mu, \tau,..., \, i = 1, 2, 3, 4,...
\end{equation}
where Greek (Latin) indexes denote flavor (mass) eigenstates. The
coupling of the HNLs to charged or neutral currents of flavor $\alpha$
is characterized by the quantities $U_{\alpha 4}$, $U_{\alpha 5}$ etc,
which we denote for convenience as $U_{\alpha}$. A generic HNL is
denoted by $\nu_h$. Its production and decay diagrams are shown in
Fig.~\ref{p:feyn}. Existing experimental results are reviewed and
discussed in Ref.~\cite{review}.

In our previous analysis~\cite{btohnl} we searched for the decays of
HNLs produced in $B$ decays. No signal was found and upper limits on
$|U_e|^2$, $|U_\mu|^2$ and $|U_e||U_\mu|$ as functions of the mass of
the HNL were set.


In this analysis, we reconstruct $\tau^- \to \pi^-_p\nu_h$ decays, where
$\pi_p$ refers to the ``prompt pion'' and the HNL decays into a
pion-lepton pair in the detector volume: $\nu_h \to \pi^\pm\ell^\mp$,
$\ell=e,\mu$ (the charge-conjugate decay mode being included throughout
this Letter).
Both combinations of $\tau$ and $\ell$ charges are retained for further
analysis. In the final state, we have two pions and a lepton: $\pi^-_p
\pi^\pm \ell^\mp$.

Following~\cite{tastet,atlas}, we interpret the result in terms of the
minimal realistic model with two quasi-degenerate HNLs with close masses
and couplings and not trivial $U_\alpha$. When two HNL masses are not
exactly the same, HNL oscillations occur and we consider two extreme
cases: the ``Dirac-like limit'', where only lepton-number
conserving final states are allowed, and the
``Majorana-like limit'', where lepton-number violating final states are
also allowed with the same branching fractions. The ratio of different
$U_\alpha$ is determined from the neutrino oscillation data. In the
normal hierarchy case (NH), the relative mixing coefficients $x_\alpha
\equiv |U_\alpha|^2/|U|^2$, $|U|^2 = \sum_\alpha |U_\alpha|^2$ $(\alpha
= e, \mu, \tau)$ are taken to be $x_e = 0.06$, $x_\mu = 0.48$ and
$x_\tau = 0.46$; for the inverted hierarchy case (IH), we use the values
$x_\alpha = 1/3$ $(\alpha = e, \mu, \tau)$~\cite{tastet}.

A distinctive feature of the HNL is its long lifetime. We can estimate
it as $c\tau \sim |U|^{-2}M(\nu_h)^{-5}$~\cite{gronau}: for
$M(\nu_h)=1\,\textrm{GeV}/c^2$ and $|U|^2=10^{-4}$, the lifetime is
$c\tau \sim 20\,\textrm{m}$; thus, a $\pi\ell$ pair forms a vertex
displaced from the interaction point (IP). BaBar~\cite{babar-lhh} and
Belle~\cite{miyazaki} previously searched for $\tau \to \ell h h'$
decays; however, both analyses required all tracks to originate in the
vicinity of the IP. For the long-lived HNL, this greatly reduces the
reconstruction efficiency. In contrast, we do not impose such a
requirement on the HNL daughters.


Results presented here are based on all available Belle data, including
on-resonance, off-resonance and energy scans. The collision energy is
around 10.58\,\textrm{GeV}. The total integrated luminosity is
$988\,\mathrm{fb}^{-1}$~\cite{lum} and the total number of
$\tau^+\tau^-$ pairs is calculated using direct production cross
sections~\cite{swagato} and $\Upsilon(NS)$ branching fractions to be
$N_{\tau\tau} = (912 \pm 13) \times 10^6$, where the error arises from
the luminosity measurement.

The Belle detector is a large-solid-angle magnetic
spectrometer that consists of a silicon vertex detector (SVD),
a 50-layer central drift chamber (CDC), an array of
aerogel threshold Cherenkov counters (ACC),  
a barrel-like arrangement of time-of-flight
scintillation counters (TOF), and an electromagnetic calorimeter
comprised of CsI(Tl) crystals (ECL) located inside 
a super-conducting solenoid coil that provides a 1.5~T
magnetic field.  An iron flux-return located outside of
the coil is instrumented to detect $K_L^0$ mesons and to identify
muons (KLM).  The detector
is described in detail elsewhere~\cite{Belle}.

To study backgrounds, we use the following Monte Carlo (MC) simulated
samples: $e^+e^- \to q\bar{q}$ $(q=u,d,s,c,b)$, $e^+e^-$, $\mu^+\mu^-$,
$\tau^+\tau^-$ and two-photon processes ($e^+e^- \to
e^+e^-\ell^+\ell^-$, $e^+e^- q\bar{q}$). Background processes are
generated by EvtGen~\cite{evtgen}, BHLUMI~\cite{bhlumi},
KKMC~\cite{kkmc}, KKMC ($\tau\tau$ production) and TAUOLA ($\tau$
decay)~\cite{tauola}, and AAFH~\cite{aafh}, respectively. We use signal
MC samples generated with different HNL masses of $M(\nu_h) =
0.2\,\textrm{GeV}/c^2$ to $1.6\,\textrm{GeV}/c^2$ (with
$0.1\,\textrm{GeV}/c^2$ step) and life times of $c\tau = 0.2, 0.5,$ and
$1.0\,\textrm{m}$ to study the response of the detector and to determine
its acceptance and signal efficiency dependence on the neutrino mass and
the distance of the decay point from the IP. This efficiency does not
depend on $c\tau$ ($|U|^2$). Signal MC samples are $\tau^+\tau^-$ pairs
where one of the $\tau$ leptons decays according to the modes under
study and the other decays generically. Signal events are generated
using EvtGen; radiative corrections are included using
PHOTOS~\cite{photos}. HNLs are produced and decayed uniformly in phase
space. GEANT3~\cite{geant3} is used to model the detector response.


Electrons are identified using the energy and shower profile in the ECL,
the light yield in the ACC and the specific ionization energy loss in
the CDC ($dE/dx$). This information is used to form an electron
($\mathcal{L}_e$) and non-electron ($\mathcal{L}_{\bar{e}}$) likelihood;
these are combined into a likelihood ratio $\mathcal{P}_{e} =
\mathcal{L}_{e}/(\mathcal{L}_e+\mathcal{L}_{\bar{e}})$~\cite{eid}.
Muons are distinguished from other charged tracks by their range and hit
profiles in the KLM. This information is utilized in a likelihood ratio
approach~\cite{muid} similar to the one used for the electron
identification.

Charged tracks with laboratory momentum greater than
$0.5\,\textrm{GeV}/c$ and electron likelihood ratio $\mathcal{P}_e >
0.9$ or muon ratio $\mathcal{P}_\mu > 0.9$ are treated as leptons. These
requirements correctly identify leptons with an efficiency of
approximately 95\% and a misidentification rate of less than 2\%. All
charged tracks not identified as leptons and satisfying the electron
veto ${\cal P}_e<0.5$ are treated as pions.

HNL candidates are formed from a pion $\pi$ and a lepton $\ell$ of the
opposite sign. The pion and lepton are then fitted to a common
vertex. HNL candidates are combined with a prompt pion $\pi_p$. The
second vertex fit of the HNL candidate and the prompt pion is performed
with a vertex constraint to the IP, which is determined run-by-run using
charged tracks. The $\chi^2$ of both vertex fits is required to be
$\chi^2/ndf<25$, where $ndf$ is the number of degrees of
freedom. Kinematics of the particles are updated after the fits are
performed. For the prompt pion, we require the closest distance to IP
along the detector symmetry axis ($dz$) to be $|dz| < 5\,\textrm{cm}$
and in the transverse plane to be $dr < 1\,\textrm{cm}$.

Since the signal $\tau$ lepton is fully reconstructed, we can utilize
the kinematic constraint of the known initial four-momentum of the
colliding $e^+e^-$ pair to define $\Delta E \equiv E(\pi_p\pi\ell) -
E_\textrm{cm}$ in the center-of-mass (c.m.) frame of a $\tau$ candidate
relative to the beam energy $E_\textrm{cm}$. In $\tau$ decays, $\Delta
E$ and $M(\pi_p\pi\ell)$ are highly correlated; therefore, we use an
elliptically shaped requirement, that encompasses $\sim$~95\% efficiency
as computed from the signal simulation.

In the rest of the event, we select tracks with $dr < 1\,\textrm{cm}$,
$|dz| < 5\,\textrm{cm}$ and a transverse momentum $p_t >
0.1\,\textrm{GeV}/c$. We classify clusters in the ECL not associated
with charged tracks as photons and require
$E(\gamma)>0.05\,\textrm{GeV}$ in the barrel ($32.2^{\circ} < \theta <
128.7^{\circ}$), $E(\gamma)>0.1\,\textrm{GeV}$ in the forward endcap
($12.4^{\circ} < \theta < 31.4^{\circ}$) and
$E(\gamma)>0.15\,\textrm{GeV}$ in the backward endcap ($130.7^{\circ}
< \theta < 155.1^{\circ}$). Events are separated into two hemispheres
by the plane perpendicular to the thrust axis
$\vec{n}_T$~\cite{thrust}, defined to maximize the thrust magnitude
value
\begin{equation}
V_T = \frac{\sum |\vec{p}_i^* \cdot \vec{n}_T|}{\sum |\vec{p}_i^*|},
\end{equation}
where $\vec{p}_i^*$ are momenta of the selected tracks, photons and
$\tau$ daughters in the c.m. frame. We require the signal hemisphere
to contain no additional tracks besides $\tau$ daughters, and the
opposite side to contain one or three tracks with a total charge
opposite that of the prompt pion.

We select well-vertexed HNL candidates using $d\phi$, the angle
between the momentum vector and decay-vertex vector of the HNL
candidate; $dz_\textrm{vtx}$, the distance between the daughter tracks
at their closest approach in the direction parallel to the beam; and
$dr$ for each track. Requirements vary depending on the presence of
SVD hits on the tracks and on the HNL candidate flight length. These
are summarized in Table~\ref{t:goodhn}. The four event types in the
Table are I: both neutrino daughter tracks have recorded hits in SVD,
II: one of the neutrino daughter tracks has recorded hits in SVD, III:
none of the neutrino daughter tracks have recorded hits in SVD, with
$M(\nu_h) \ge 0.8\,\textrm{GeV}/c^2$, and IV: no SVD hits and
$M(\nu_h)<0.8\,\textrm{GeV}/c^2$. There is a large contamination from
the conversion photons $\gamma \to ee$ in the last category.

\begin{table}[htbp]
  \caption{Summary of the vertex requirements.}
  \label{t:goodhn}
  \begin{center}
    \begin{tabular}{c|c|c|c}
      Type & $d\phi$, rad & $z_\mathrm{vtx}$, cm & $dr$, cm \\
      \hline \hline
      I   & $<0.02$  & $<0.06$ & $>0.07$ \\
      II  & $<0.024$ & $<1.5$  & $>0.08$ \\
      III & $<0.16$  & $<3.0$  & $>0.1$ \\
      IV  & $<0.16$  & $<3.0$  & $>1.0$ \\
    \end{tabular}
  \end{center}
\end{table}

\begin{figure}[htbp]
  \includegraphics[width=0.48\textwidth,clip=true]{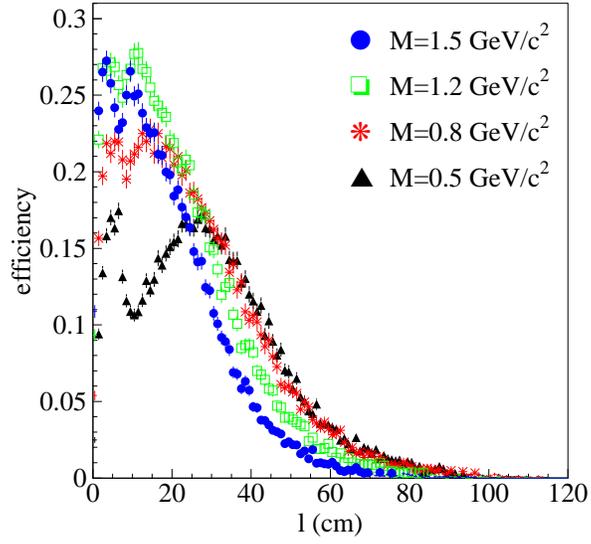}
  \caption{Dependence of the HNL reconstruction efficiency on the
    neutrino travel distance $l$ for different neutrino masses
    $M(\nu_h)$. Efficiency is almost identical for $e$ and $\mu$.}
  \label{p:eff}
\end{figure}

Figure~\ref{p:eff} shows the efficiency of HNL reconstruction with all
requirements applied as a function of the reconstructed travel distance
$l$ for several mass hypotheses.


The number of neutrinos detected by this method is (in units where
$\hbar=c=1$)~\cite{suppl}
\begin{widetext}
  \begin{align}
    \label{e:nnu}
n(\nu_h) & = 2N_{\tau\tau}\ \mathcal{B}(\tau \to \pi\nu_h)\ \mathcal{B}(\nu_h \to \pi\ell)\ \frac{m\Gamma}{p} \int \exp\Big(-\frac{m\Gamma l}{p}\Big) \varepsilon(m, l) dl\nonumber\\
         & = |U_\tau|^2|U_\ell|^2\ 2N_{\tau\tau}\ f_1(m)\ f_2(m)\ \frac{m}{p} \int \exp\Big(-\frac{m\Gamma l}{p}\Big) \varepsilon(m, l) dl,
  \end{align}
\end{widetext}
where $N_{\tau\tau}$ is the number of $\tau\tau$ pairs,
$\mathcal{B}(\tau \to \pi\nu_h)$ is the branching fraction for $\nu_h$
production, $\mathcal{B}(\nu_h \to \pi\ell)$ is the branching fraction
of the reconstructed decay, $m$, $p$ and $\Gamma = \Gamma(m, U)$ are the
mass, momentum and full width of the HNL, respectively, and
$\varepsilon(m, l)$ is the reconstruction efficiency of the HNL of mass
$m$ decaying at a distance $l$ from the IP. The momentum $p$ is
approximated by the mean value for a given mass, determined from the
signal MC simulation. To factor out the $|U_\ell|^2$ dependence, we
define functions $f_{1,2}(m)$ as $|U_\tau|^2f_1(m) \equiv
\mathcal{B}(\tau \to \pi\nu_h)$ and $|U_\ell|^2f_2(m) \equiv
\Gamma(\nu_h \to \pi\ell) = \mathcal{B}(\nu_h \to \pi\ell) \Gamma$,
where $\ell$ denotes the flavor ($e,\ \mu$) of the charged lepton
produced in the $\nu_h$ decay. Integration is performed over the full
volume used to reconstruct the HNL vertex. The expressions for
$\mathcal{B}(\tau \to \pi \nu_h)$, $\Gamma(\nu_h \to \pi\ell)$ and the
full neutrino width $\Gamma$ are taken from Ref.~\cite{gorbunov} and
require only general assumptions (i.e., they are not specific to the
$\nu$MSM model). In the Majorana case $\Gamma(\nu_h \to \pi\ell)$ is
twice that in the Dirac case.
Given number of observed events, we solve Eq.~\ref{e:nnu} for the
variable $|U|^2$ using the relative mixing coefficients $x_\alpha$
defined above.


The systematic uncertainties in number of events calculated according to
Eq.~\ref{e:nnu} are enumerated in Table~\ref{t:syserr}. We estimate the
systematic uncertainties of event selection criteria from the
differences in their efficiencies obtained in data and MC
simulation. Since all particles used in the systematic uncertainty study
decay relatively close to the IP compared to the expectation for an HNL,
we require where possible that the decay vertices be farther than 4\,cm
from the IP in the transverse plane to put more weight on large decay
lengths. To estimate the systematic uncertainty due to tracking, we
compare the number of fully and partially reconstructed $D^{*+}$ decays
in the decay chain $D^{*+} \to D^0 \pi^+$, $D^0 \to K_S^0 \pi^+ \pi^-$,
$K_S^0 \to \pi^+\pi^-$, where in the latter case one of the pions from
the $K_S^0$ is explicitly left unreconstructed. To estimate the
systematic uncertainty of the lepton identification, we reconstruct
$J/\psi\to\ell^+\ell^-$, $\ell = e, \mu$ events, where one of the
daughter particles is identified as an electron or muon. The difference
of the identification efficiency of the other daughter between data and
MC simulation is treated as a systematic uncertainty. To estimate the
systematic uncertainty of the vertex quality requirements we apply them
to $K^0_S$ decays, which have a topology similar to HNL decays. Signal
events were generated using EvtGen, which is not optimized for $\tau$
decays. To estimate the effect of this we prepared two samples of
$\tau\tau$ events --- one generated with EvtGen and the other with KKMC
and TAUOLA --- then reconstructed $\tau \to \ell \nu \nu$ decays, where
$\ell = e, \mu$, using the same tagging as for the signal events and
compared reconstruction efficiency in both cases.
The phase
space model may not give the correct angle distribution of the HNL. We
vary it by reweighting generated events and treat the change as a
systematic uncertainty.
The calculation uncertainty comes from the
efficiency and momentum approximations in Eq.~\ref{e:nnu} and was
estimated by comparing predicted and observed number of events in
different subsets of the signal MC simulation. Systematic uncertainties
induced by the fitting procedure were found by varying the signal
resolution and background shape within their errors. The theoretical
uncertainty arises from uncertainties in the constants used in
Eq.~\ref{e:nnu}. Correlations between different systematic uncertainties
are found to be small and are ignored. All systematic uncertainties are
summed in quadrature, leading to total systematic uncertainties of 16\%
and 12\% for the $\pi\pi e$ and $\pi\pi\mu$ modes, respectively. The
largest contributions are lepton identification (12\% and 6\% for the
electron and muon identification, respectively) and vertex quality
requirements (5.3\%).

\begin{table}[htp]
\caption{Summary of systematic uncertainties in number of events
  calculated according to Eq.~\ref{e:nnu}.}
\label{t:syserr}
\begin{center}
\begin{tabular}{c|c}
Requirement & Systematic uncertainty, \% \\
\hline \hline
Tracking                      & 1.2 \\
${\cal P}_e(\ell)$            & 12 \\
${\cal P}_\mu(\ell)$          & 6 \\
Vertex quality                & 5.3 \\
$N_{\tau\tau}$                  & 1.4 \\
Generator                     & 2 \\
Angle distribution            & 5 \\
Calculation                   & 4 \\
Signal resolution             & 2.5 \\
Background shape              & 4 \\
Theoretical                   & 0.35 \\
\hline
Total, $e/\mu$                & 16/12 \\
\end{tabular}
\end{center}
\end{table}


We study the dependence of the coupling constant $|U|^2$ on the HNL mass
using simultaneous fit to $\pi\pi e$ and $\pi\pi\mu$ modes taking into
account the relative mixing coefficients $x_\alpha$ defined above.

\begin{figure*}[tp]
    \begin{tabular}{cc}
       \includegraphics[width=0.48\textwidth,clip=true]{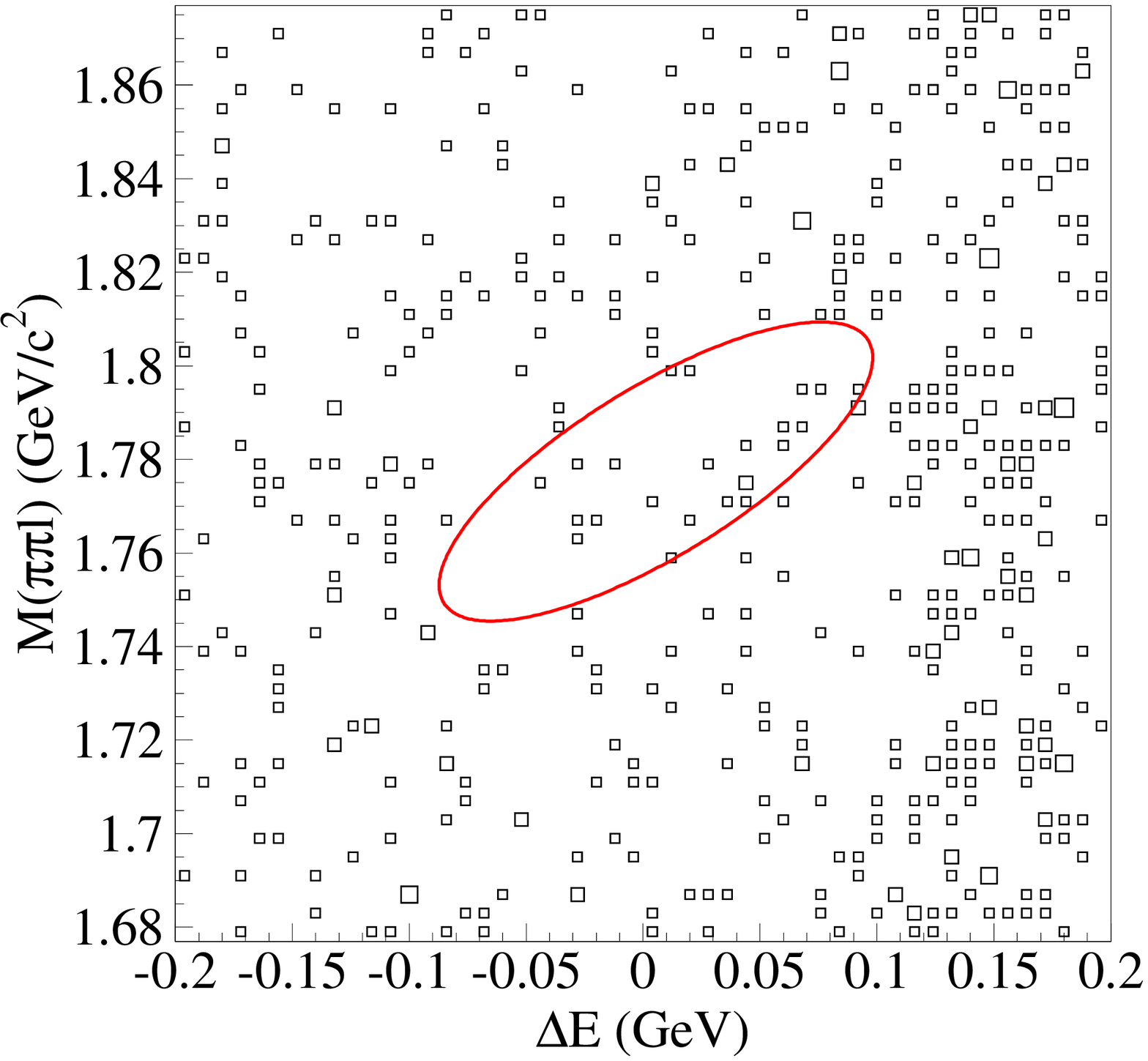}&
       \includegraphics[width=0.48\textwidth,clip=true]{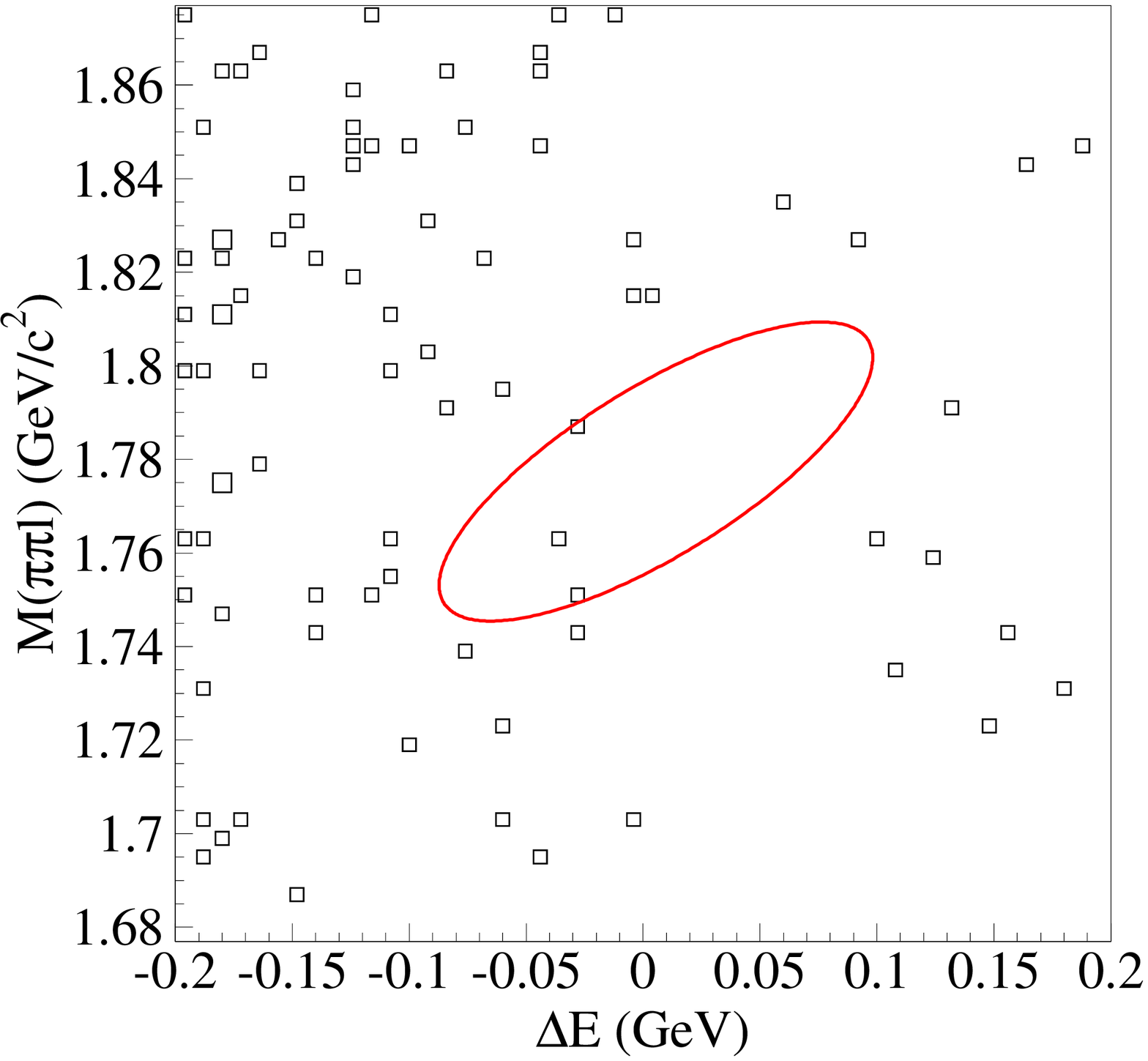}\\
       \textit{a)} & \textit{b)}\\
    \end{tabular}
    \caption{$\Delta E$ vs $M(\pi\pi\ell)$ distributions with all
      requirements but $\Delta E$ and $M(\pi\pi\ell)$ imposed for
      $\pi\pi e$~\textit{(a)} and $\pi\pi\mu$~\textit{(b)} in data. The
      signal region is shown as a red ellipse.}
    \label{p:de-vs-mtau_data}
\end{figure*}

\begin{figure}[htbp]
  \includegraphics[width=0.48\textwidth,clip=true]{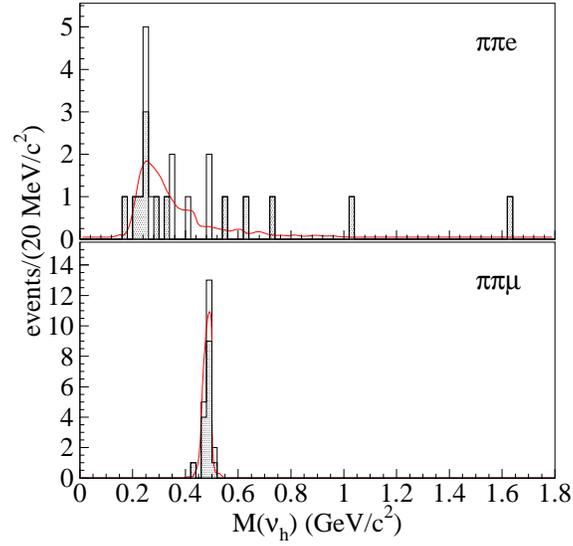}
  \caption{Final distributions of $M(\nu_h)$ for $\pi\pi e$ and
    $\pi\pi\mu$ reconstruction modes in data. The filled histograms
    are for candidates with opposite-charge $\tau$ and $\ell$, while
    the open histograms are for candidates with same-charge
    combinations. The curves are the fits with the signal yield
    fixed at zero.}
  \label{p:result_num}
\end{figure}

The $\Delta E$ vs $M(\pi\pi\ell)$ distributions with all requirements
but $\Delta E$ and $M(\pi\pi\ell)$ imposed are shown in
Fig.~\ref{p:de-vs-mtau_data}. The mass distributions after application
of all reconstruction requirements are shown in Fig.~\ref{p:result_num}
for the same-charge $\tau$ and $\ell$ (``Dirac-like limit'') and both
same- and opposite-charge combinations (``Majorana-like limit''). From
the background MC simulation study, we expect to see a wide peak around
$\sim 0.2\,\textrm{GeV}/c^2$ from the conversion process $\gamma \to
e^+e^-$ in the $\pi\pi e$ mode and a narrow peak from the $K_S^0 \to
\pi^+\pi^-$ process at $\sim 0.48\,\textrm{GeV}/c^2$ in the $\pi\pi\mu$
mode. Since the conversion peak is wide, we can distinguish a narrow
signal peak from the HNL decay under it, but since the $K_S^0$ peak is
narrow, we exclude the $K_S^0$ region at $0.464-0.494\,\textrm{GeV}/c^2$
from consideration, which corresponds to $\pm 2\sigma$ of the peak
width.

The HNL mass is unknown and we search for it in the kinematically
accessible region for the mass; for the decays under study, this lies
between $M_\pi+M_\ell$ and $M_\tau-M_\pi$. We perform a series of binned
likelihood fits to the mass distributions using the sum of a Gaussian
signal function and background (described below) varying the mass
hypothesis in each fit. The neutrino mass is set to the center of a
histogram bin which has a width of $2\,\textrm{MeV}/c^2$. The
signal-shape parameters used in the fits to data are fixed to those
obtained by fitting simulated events. The width evolves linearly from
$\sim 3\,\textrm{MeV}/c^2$ for $M(\nu_h)=0.2\,\textrm{GeV}/c^2$ to
$\sim 10\,\textrm{MeV}/c^2$ for $M(\nu_h)=1.6\,\textrm{GeV}/c^2$. The
background is described by the sum of a constant and the conversion peak
described above in the $\pi\pi e$ subset and by a constant and the
$K_S^0$ peak in the $\pi\pi\mu$ subset. The functions for the peaking
components are defined as smoothed histograms from the background MC
simulation. Yields of all components are free parameters of the fit.

\begin{figure}[htbp]
  \includegraphics[width=0.48\textwidth,clip=true]{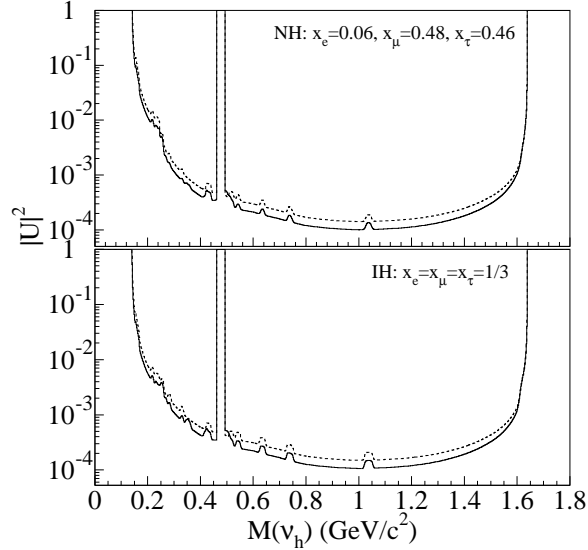}
  \caption{Upper limits at 95\%~CL on $|U|^2$. The upper (lower) plot is
    for the normal (inverted) hierarchy. The solid (dashed) curve shows
    the result in the ``Majorana-like limit'' (the ``Dirac-like
    limit''). The excluded area is above the curves.}
  \label{p:result_upli}
\end{figure}

The statistical significance of the HNL signal is defined as
$\sqrt{-2\ln{L_0}/{L}}$, where $L_0$ and $L$ are the likelihoods
returned by the fit with the signal yield fixed at zero and at the
fitted value, respectively. The maximum local significance
in four fits does not exceed $2.5\,\sigma$,
and we set upper limits on $|U|^2$ at 95\%~CL in the ``Dirac-like
limit'' and the ``Majorana-like limit'' for the two neutrino-mass
hierarchy scenarios bin by bin using \texttt{pyhf}
package~\cite{pyhf-git,pyhf-paper}.
The resulting upper limits on the coupling constants at 95\%~CL are
shown in Fig.~\ref{p:result_upli}. Comparison of all four upper limits
on one plot may be found in the Supplimentary Materials~\cite{suppl}.


In conclusion, we search for a heavy neutrino in $\tau$ decays and
observe no significant signal. This is the first study of a mixed
couplings of heavy neutrinos to $\tau$ leptons and light-flavor
leptons. Upper limits on the mixing of HNLs to the active neutrinos in
the mass range $0.2-1.6\,\textrm{GeV}/c^2$ are set. The maximum
sensitivities are achieved around $1.0\,\textrm{GeV}/c^2$ and the
corresponding upper limits at 95\%~CL for $|U|^2$ are $1.4 \times
10^{-4}$ ($1.5 \times 10^{-4}$) in the ``Dirac-like limit'' for the
normal (inverted) hierarchy and $1.0 \times 10^{-4}$ ($1.1 \times
10^{-4}$) in the ``Majorana-like limit'' for the normal (inverted)
hierarchy.




This work, based on data collected using the Belle detector, which was
operated until June 2010, was supported by 
the Ministry of Education, Culture, Sports, Science, and
Technology (MEXT) of Japan, the Japan Society for the 
Promotion of Science (JSPS), and the Tau-Lepton Physics 
Research Center of Nagoya University; 
the Australian Research Council including grants
DP180102629, 
DP170102389, 
DP170102204, 
DE220100462, 
DP150103061, 
FT130100303; 
Austrian Federal Ministry of Education, Science and Research (FWF) and
FWF Austrian Science Fund No.~P~31361-N36;
the National Natural Science Foundation of China under Contracts
No.~11675166,  
No.~11705209;  
No.~11975076;  
No.~12135005;  
No.~12175041;  
No.~12161141008; 
Key Research Program of Frontier Sciences, Chinese Academy of Sciences (CAS), Grant No.~QYZDJ-SSW-SLH011; 
the Ministry of Education, Youth and Sports of the Czech
Republic under Contract No.~LTT17020;
the Czech Science Foundation Grant No. 22-18469S;
Horizon 2020 ERC Advanced Grant No.~884719 and ERC Starting Grant No.~947006 ``InterLeptons'' (European Union);
the Carl Zeiss Foundation, the Deutsche Forschungsgemeinschaft, the
Excellence Cluster Universe, and the VolkswagenStiftung;
the Department of Atomic Energy (Project Identification No. RTI 4002) and the Department of Science and Technology of India; 
the Istituto Nazionale di Fisica Nucleare of Italy; 
National Research Foundation (NRF) of Korea Grant
Nos.~2016R1\-D1A1B\-02012900, 2018R1\-A2B\-3003643,
2018R1\-A6A1A\-06024970, RS\-2022\-00197659,
2019R1\-I1A3A\-01058933, 2021R1\-A6A1A\-03043957,
2021R1\-F1A\-1060423, 2021R1\-F1A\-1064008, 2022R1\-A2C\-1003993;
Radiation Science Research Institute, Foreign Large-size Research Facility Application Supporting project, the Global Science Experimental Data Hub Center of the Korea Institute of Science and Technology Information and KREONET/GLORIAD;
the Polish Ministry of Science and Higher Education and 
the National Science Center;
the Ministry of Science and Higher Education of the Russian Federation, Agreement 14.W03.31.0026, 
and the HSE University Basic Research Program, Moscow; 
University of Tabuk research grants
S-1440-0321, S-0256-1438, and S-0280-1439 (Saudi Arabia);
the Slovenian Research Agency Grant Nos. J1-9124 and P1-0135;
Ikerbasque, Basque Foundation for Science, Spain;
the Swiss National Science Foundation; 
the Ministry of Education and the Ministry of Science and Technology of Taiwan;
and the United States Department of Energy and the National Science Foundation.
These acknowledgements are not to be interpreted as an endorsement of any
statement made by any of our institutes, funding agencies, governments, or
their representatives.
We thank the KEKB group for the excellent operation of the
accelerator; the KEK cryogenics group for the efficient
operation of the solenoid; and the KEK computer group and the Pacific Northwest National
Laboratory (PNNL) Environmental Molecular Sciences Laboratory (EMSL)
computing group for strong computing support; and the National
Institute of Informatics, and Science Information NETwork 6 (SINET6) for
valuable network support.

\clearpage

\newpage

\appendix
\section{Derivation of Eq. 3}
\label{suppl}

If a particle with a mass $m$ and width $\Gamma$ has a momentum $p$,
then the probability that it travels distance $l$ or greater is (in units where
$\hbar=c=1$)
\begin{equation}
P(l)=\exp\Big(-\frac{m\Gamma l}{p}\Big)\nonumber,
\end{equation}
thus probability to decay in a segment $dl$ at distance $l$ is
\begin{equation}
dP(l)= \frac{m\Gamma}{p} \exp\Big(-\frac{m\Gamma l}{p}\Big)dl\nonumber.
\end{equation}
The number of neutrinos detected in the Belle detector is
\begin{widetext}
  \begin{align}
        n(\nu_h) & = N_0 \int \varepsilon(m, l) dP(l)\nonumber\\
         & = 2N_{\tau\tau}\ \mathcal{B}(\tau \to \pi\nu_h)\ \mathcal{B}(\nu_h \to \pi\ell)\ \frac{m\Gamma}{p} \int \exp\Big(-\frac{m\Gamma l}{p}\Big) \varepsilon(m, l) dl\nonumber\\
         & = |U_\tau|^2|U_\ell|^2\ 2N_{\tau\tau}\ f_1(m)\ f_2(m)\ \frac{m}{p} \int \exp\Big(-\frac{m\Gamma l}{p}\Big) \varepsilon(m, l) dl\nonumber\\
         & = |U|^2\ x_{\tau}x_{\ell}\ 2N_{\tau\tau}\ f_1(m)\ f_2(m)\ \frac{m}{p} \int \exp\Big(-\frac{m\Gamma l}{p}\Big) \varepsilon(m, l) dl,
  \end{align}
\end{widetext}
where $N_0 = 2N_{\tau\tau}\ \mathcal{B}(\tau \to
\pi\nu_h)\ \mathcal{B}(\nu_h \to \pi\ell)$ is the total number of
$\pi\ell$ pairs produced in the $\tau \to \pi\nu_h$, $\nu_h \to \pi\ell$
decay chain and the rest of definitions is the same as in the paper.

The $|U_\tau|^2$ coupling comes from the branching fraction $\mathcal{B}(\tau
\to \pi\nu_h)$, and the $|U_\ell|^2$ coupling comes from the partial width
$\Gamma(\nu_h \to \pi\ell)=\mathcal{B}(\nu_h \to \pi\ell)\Gamma$,
$|U_\alpha|^2=\ x_{\alpha}|U|^2$. The full width $\Gamma$ is calculated by
summing over a number of two- and three-body decays taking into account the
relative mixing coefficients $x_\alpha$.

\begin{figure*}[htbp]
  \includegraphics[width=0.98\textwidth,clip=true]{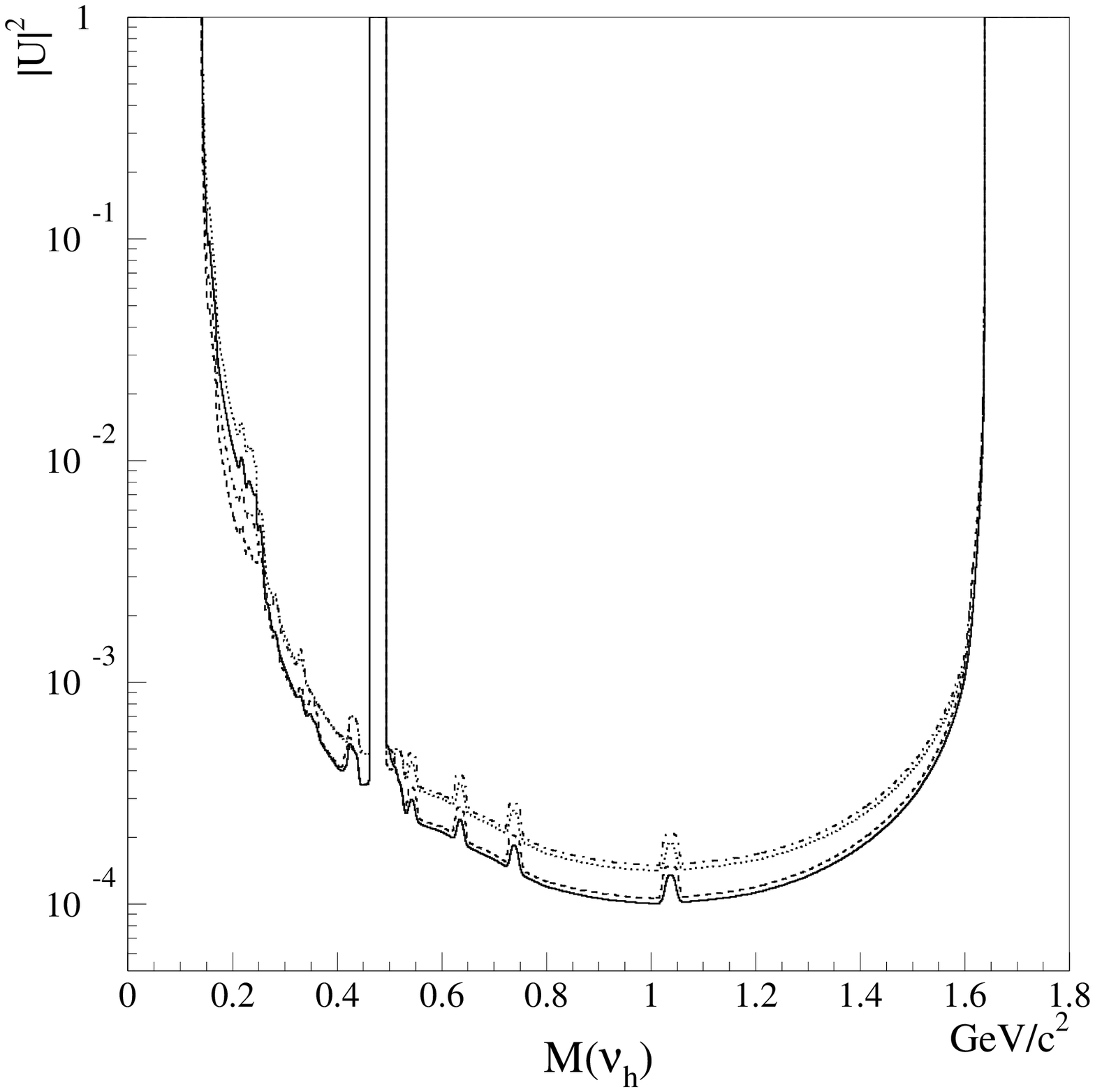}
  \caption{Upper limits at 95\%~CL on $|U|^2$. The solid, dashed, dotted
    and dash-dotted curves show the result in the ``Majorana-like
    limit'' for normal ($x_e = 0.06$, $x_\mu = 0.48$ and $x_\tau =
    0.46$) and inverted ($x_\alpha = 1/3, \alpha = e, \mu, \tau$)
    hierarchy and in the ``Dirac-like limit'' for normal and inverted
    hierarchy, respectively. The excluded area is above the curves.}
  \label{p:result_upli_full}
\end{figure*}


\begin{thebibliography}{99}

\bibitem{pdg}
  R.~L.~Workman \textit{et al.} (Particle Data Group), Prog. Theor. Exp. Phys. \textbf{2022}, Issue 8, 083C01 (2022).

\bibitem{katrin}
  M.~Aker \textit{et al.} (KATRIN Collaboration), Nat. Phys. \textbf{18}, 160 (2022) [arXiv:2105.08533].

\bibitem{seesaw}
  P. Minkowski, Phys. Lett. \textbf{B 67}, 421 (1977);\\
  T. Yanagida, in \textit{Proc. of the Workshop on Grand Unified Theory and Baryon Number of the Universe}, KEK, Japan, 1979;\\
  M. Gell-Mann, P. Ramond and R. Slansky, (1980), Print-80-0576 (CERN);\\
  R.N. Mohapatra and G. Senjanovic, Phys. Rev. Lett. \textbf{44}, 912 (1980).

\bibitem{numsm}
  T. Asaka, S. Blanchet and M. Shaposhnikov, Phys. Lett. \textbf{B 631}, 151 (2005) [arXiv:hep-ph/0503065].

\bibitem{review}
  F.~F.~Deppisch, P.~S.~Bhupal~Dev, A.~Pilaftsis, New J. Phys. \textbf{17}, 075019 (2015) [arXiv:1502.06541];\\
  P.~D.~Bolton, F.~F.~Deppisch, P.~S.~Bhupal~Dev, J. High Energ. Phys. \textbf{03}, 170 (2020) [arXiv:1912.03058].

\bibitem{btohnl}
  D.~Liventsev \textit{et al.} (Belle Collaboration), Phys. Rev. \textbf{D. 87}, 071102 (2013) [arXiv:1301.1105].

\bibitem{tastet}
  J.-L.~Tastet, O.~Ruchayskiy, and I.~Timiryasov, J. High Energ. Phys. \textbf{12} 182 (2021), [arXiv:2107.12980].

\bibitem{atlas}
  ATLAS collaboration, arXiv:2204.11988.

\bibitem{gronau}
  M.~Gronau, C.~N.~Leung and J.~L.~Rosner, Phys. Rev. \textbf{D 29}, 2539 (1984).

\bibitem{babar-lhh}
  B.~Aubert \textit{et al.} (BaBaR Collaboration), Phys. Rev. Lett. \textbf{95}, 191801 (2005) [arXiv:hep-ex/0506066].

\bibitem{miyazaki}
  Y.~Miyazaki \textit{et al.} (Belle Collaboration), Phys. Lett. \textbf{B 719}, 346 (2013) [arXiv:1206.5595].

\bibitem{lum}
  J.~Brodzicka \textit{et al.} (Belle Collaboration), Prog. Theor. Exp. Phys. \textbf{2012}, Issue 1, 4D001 (2012) [arXiv:1212.5342].

\bibitem{swagato}
  S.~Banerjee, B.~Pietrzyk, J.~M.~Roney, and Z.~Was, Phys. Rev. \textbf{D 77}, 054012 (2008).

\bibitem{Belle}
  A.~Abashian {\it et al.} (Belle Collab.), Nucl. Instr. and Meth. A \textbf{479}, 117 (2002).

\bibitem{evtgen}
  D. J. Lange, Nucl. Instr. and Meth. A \textbf{462}, 152 (2001).

\bibitem{kkmc}
  S.~Jadach, B.~Ward, and Z.~W\c{a}s, Comput. Phys. Commun. \textbf{130}, 260 (2000).

\bibitem{tauola}
  N.~Davidson \textit{et al.}, Comput. Phys. Commun. \textbf{183}, 821 (2012).

\bibitem{bhlumi}
  S.~Jadach \textit{et al.}, Comput. Phys. Commun. \textbf{70}, 305 (1992).

\bibitem{aafh}
  F.~A.~Berends \textit{et al.}, Comput. Phys. Commun. \textbf{40}, 285 (1986).

\bibitem{photos}
  E.~Barberio and Z.~Was, Comput. Phys. Commun. \textbf{79}, 291 (1994).

\bibitem{geant3}
  R.~Brun \textit{et al.}, CERN Report No. DD/EE/84-1 (1984).

\bibitem{eid}
  K.~Hanagaki \textit{et al.}, Nucl. Instr. and Meth. A \textbf{485}, 490 (2002).

\bibitem{muid}
  A.~Abashian \textit{et al.}, Nucl. Instr. and Meth. A \textbf{491}, 69 (2002).

\bibitem{thrust}
  S.~Brandt, C.~Peyrou, R.~Sosnowski, and A.~Wroblewski, Phys. Lett. \textbf{12}, 57 (1964);\\
  E.~Farhi, Phys. Rev. Lett. \textbf{39}, 1587 (1977).

\bibitem{suppl}
	See Supplimentary material (Appendix~\ref{suppl}).

\bibitem{gorbunov}
  K.~Bondarenko, A.~Boyarsky, D.~Gorbunov, and O.~Ruchayskiy, J. High Ener. Phys. \textbf{11}, (2018) 032,
  [arXiv:1805.08567].


\bibitem{pyhf-git}
  L.~Heinrich, M.~Feickert, and G.~Stark, ``pyhf:v0.7.2'',
https://github.com/scikit-hep/pyhf/releases/tag/v0.7.2, doi: 10.5281/zenodo.1169739.

\bibitem{pyhf-paper}
  L.~Heinrich \textit{et al.}, J. of Open Source Software \textbf{6(58)}, 2823 (2021), https://doi.org/10.21105/joss.02823.

\end{thebibliography}
\end{document}